\documentclass[aps,prl,reprint,groupedaddress,floatfix]{revtex4-2}
\usepackage{graphicx}
\usepackage{amssymb}
\usepackage{xcolor,soul}
\begin{document}

\title{Transport-Enabled Entangling Gate for Trapped Ions}

\author{Holly N. Tinkey}
\email{holly.tinkey@gtri.gatech.edu}
\author{Craig R. Clark}
\author{Brian C. Sawyer}
\author{Kenton R. Brown}

\affiliation{Georgia Tech Research Institute, Atlanta, Georgia 30332, USA}

\begin{abstract}
We implement a 2-qubit entangling M{\o}lmer-S{\o}rensen interaction by transporting two cotrapped $^{40}\mathrm{Ca}^{+}$ ions through a stationary, bichromatic optical beam within a surface-electrode Paul trap. We describe a procedure for achieving a constant Doppler shift during the transport which uses fine temporal adjustment of the moving confinement potential. The fixed interaction duration of the ions transported through the laser beam as well as the dynamically changing ac Stark shift require alterations to the calibration procedures used for a stationary gate. We use the interaction to produce Bell states with fidelities commensurate to those of stationary gates performed in the same system. This result establishes the feasibility of actively incorporating ion transport into quantum information entangling operations.
\end{abstract}

\maketitle

Recent progress in the ability to control trapped-ion positions and velocities offers opportunities to explore novel roles for active transport in quantum logic operations. Transport operations are essential to most modern ion trap experiments to enable loading, individual detection, and individual addressing \cite{kielpinski_architecture_2002,pino_demonstration_2021}. Maturing trap design and electrical potential control hardware have led to impressive feats of fast shuttling \cite{walther_controlling_2012, blakestad_near-ground-state_2011}, fast ion separation \cite{bowler_coherent_2012, ruster_experimental_2014}, optical phase control \cite{tinkey_quantum_2021}, junction transport \cite{blakestad_near-ground-state_2011, wright_reliable_2013}, and ion chain rotation \cite{van_mourik_coherent_2020, splatt_deterministic_2009}.

An architecture that incorporates ion transport directly into quantum gates was proposed as an approach to reduce overall optical power and timing precision requirements within large trap arrays \cite{leibfried_transport_2007}. This architecture was partially realized by de Clercq et al., who performed 1-qubit operations in parallel on two $^{9}\mathrm{Be}^{+}$ ions by transporting them through reflected, co-propagating beams in separate trap regions \cite{de_clercq_parallel_2016}. This work did not demonstrate the 2-qubit entangling operations that would be necessary for a universal gate set \cite{divincenzo_two-bit_1995}.

Here we demonstrate 2-qubit entangling gates performed on trapped ions during transport. In contrast to the hyperfine qubits envisioned in ref.~\cite{leibfried_transport_2007}, we address an optical qubit transition between an electronic ground state and a metastable excited state of $^{40}$Ca$^+$ ions confined within a surface Paul trap, where we can perform both one-qubit and two-qubit gates with a single, global beam. We modify the time dependence of the transport potential to control the velocity of the ions across an 80~$\mu$m trap region, and we apply a bichromatic field during this transit to produce a M{\o}lmer-S{\o}rensen (MS) entangling interaction \cite{sorensen_quantum_1999}. To compensate for the time-varying ac Stark shift that the ions experience as they traverse the optical beam, we leverage small changes in the Doppler shift and thereby obtain fidelities tantamount to those of stationary gates in the system.

In these experiments, two $^{40}$Ca$^+$ ions are confined 58~$\mu$m above a surface-electrode linear Paul trap \cite{shappert_spatially_2013}. A radio-frequency potential with 176~V amplitude at 56.4~MHz applied to long electrodes on both sides of the trap axis provides radial confinement. Arbitrary waveform generators (AWGs) deliver potentials with a maximum amplitude of $\pm 12$~V and a 5~ns sampling rate to 42 electrodes to control the strength and location of the axial potential minimum~\cite{tinkey_quantum_2021}. The center-of-mass (c.m.) and breathing-motion (BM) axial mode frequencies are $\omega_\mathrm{c.m.}/(2\pi) = 1.41$~MHz and $\omega_\mathrm{BM}/(2\pi)  =2.45$~MHz \cite{wineland_experimental_1998}. We use the ground $|S_{1/2}, m_j = -1/2\rangle$ ($|S\rangle$) and metastable $|D_{5/2}, m_j = -1/2\rangle$ ($|D\rangle$) electronic states of the ions as our qubit, with $m_j$ denoting the angular momentum projection of each state; we coherently manipulate the populations in these states with a narrow-linewidth 729~nm laser beam oriented at $45^\circ$ to the trap axis and with a waist of 15~$\mu$m. We distinguish populations in the two-ion bright state $P_2$ ($|SS\rangle$), the one-ion bright subspace $P_1$ (superpositions of $|SD\rangle$ and $|DS\rangle$), and the dark state $P_0$ ($|DD\rangle$) using fluorescence detected by a single photomultiplier tube while illuminating the ions with 397~nm and 866~nm light \cite{roos_quantum_1999}. The trap is housed in a room temperature ultrahigh vacuum chamber with windows for optical access, and the chamber and beam-delivery optics are surrounded by a mu-metal enclosure.

\begin{figure}[htpb]
\includegraphics[height=75mm,width=0.45\textwidth]{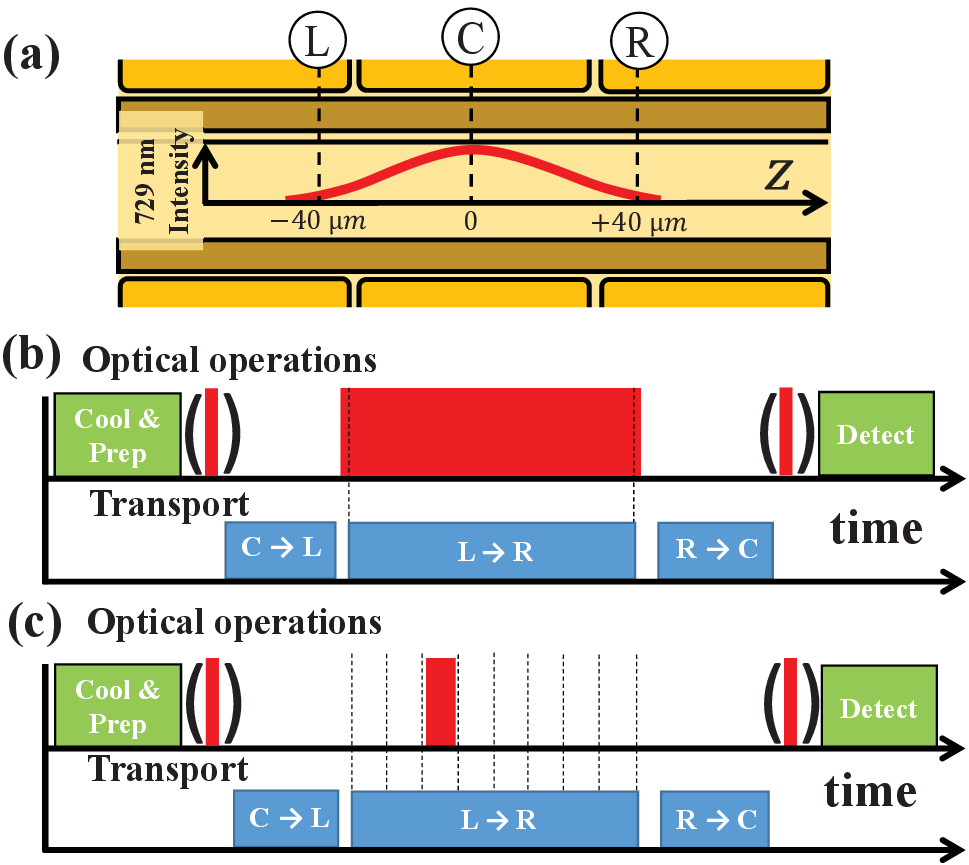}
\caption{\label{fig:Fig1} (a) Diagram of the trap and 729~nm beam intensity profile along the trap axis with rf electrodes in dark gold parallel to the trap axis and segmented electrodes above and below them. Ions begin in the center of a 60 $\mu$m wide electrode at position C. L and R denote the leftmost and rightmost portions of the transport region. (b) Experimental sequence for full transport measurements. Green blocks indicate optical operations (cooling, state preparation) before and after transport, red blocks indicate 729~nm pulses, and blue blocks indicate transport operations between indicated positions. (c) Segmented measurement sequence in which the 729~nm beam is pulsed only during a 20~$\mu$s segment of the full transport duration. Red pulses in parentheses indicate 729~nm pulses used in parity analysis and red-sideband spectroscopy measurements.}
\end{figure}

Figure~\ref{fig:Fig1} illustrates the trap geometry and experimental sequences. The ions first undergo Doppler cooling, sideband cooling (axial c.m. and BM modes), and state preparation at position C and are transported 40~$\mu$m to the left (position L) in 25~$\mu$s. The entangling interaction is then applied during an 80~$\mu$m transport (L to R) at 0.5~m/s ( 160~$\mu$s duration). For our beam geometry, this motion Doppler shifts the 729~nm beam tones by 0.5~MHz. The ions then are returned to C in 10~$\mu$s for additional 729~nm pulses and final state detection. For interactions that take place during the entire L$\rightarrow$R transport duration, the 729~nm beam is switched on 2~$\mu$s before the transport starts and off 2~$\mu$s after the transport has finished [Fig.~\ref{fig:Fig1}(b)]\footnote{The illumination of stationary ions before and after transport off-resonantly drives the carrier transition only weakly due to the Doppler shift and the reduced intensity at the edges of transport. We estimate an upper bound of $6\times10^{-5}$ for the error contribution to the gate.}. For detailed in-flight spectroscopy experiments, we divide the transport into eight portions and probe these segments using 729~nm pulses with 20~$\mu$s duration each [Fig.~\ref{fig:Fig1}(c)]. 

Our MS entangling interaction requires two optical tones applied simultaneously with frequencies detuned near the red and blue BM sidebands \cite{sorensen_entanglement_2000}; we choose the BM mode because it has a low heating rate and is not excited strongly during linear transport. We create this bichromatic field by passing the 729~nm beam through an acousto-optic modulator with two rf tones applied; both diffracted output beams are coupled into a single optical fiber to deliver co-propagating laser fields to the trapped ions \cite{akerman_universal_2015, benhelm_towards_2008}). For an interaction duration $\tau$, the optical tone frequencies and powers are chosen such that the gate transforms ions initialized in the ground state $|SS\rangle$ into a maximally entangled Bell state $(|SS\rangle - i |DD\rangle)/\sqrt{2}$. We quantify the fidelity $F$ of the state produced in this way using a combination of two experimental results: (1) the populations $P_0$ and $P_2$ after the gate and (2) the amplitude $A$ of a parity signal constructed by applying a global ${\pi/2}$ pulse with varying phase after the gate \cite{leibfried_experimental_2003}. 

We determine several initial sets of dc electrode voltages to confine the ions statically at 2~$\mu$m intervals along the trap axis by creating a three-dimensional discretized model of the ideal trap and solving for potentials that produce harmonic confinement \cite{charles_doret_controlling_2012}. For simple shuttling operations, linear interpolation between these harmonic potentials is sufficient to transport ions from one position to another. We use this preliminary sequence of voltages, called a waveform, as a starting point for transport across the 80~$\mu$m experimental region. Unfortunately, the confining fields produced in practice can differ from the ideal solutions due to trap fabrication imperfections, stray fields, and dc electrode filter distortions, so we calibrate modifications to the waveform to correct for variations in axial confinement strength and transport velocity.

Because the entangling operation relies on interactions with an axial motional mode, producing constant axial confinement during transport simplifies the gate implementation. For this we take measurements of the c.m. mode frequency of an ion in a stationary potential every 5~$\mu$m along the transport region: we apply a rf excitation to a nearby electrode and measure the detected fluorescence as the excitation frequency is varied.
These measurements reveal deviations in the harmonic potential strength at different locations in the transport region (as high as 4.6\% variation in $\omega_\mathrm{c.m.}$), which we correct with a multiplicative scaling on all of the electrode voltages responsible for harmonic confinement, but not on those compensating for stray electric fields.

\begin{figure}[htpb]
\includegraphics[height=80mm,width=0.45\textwidth]{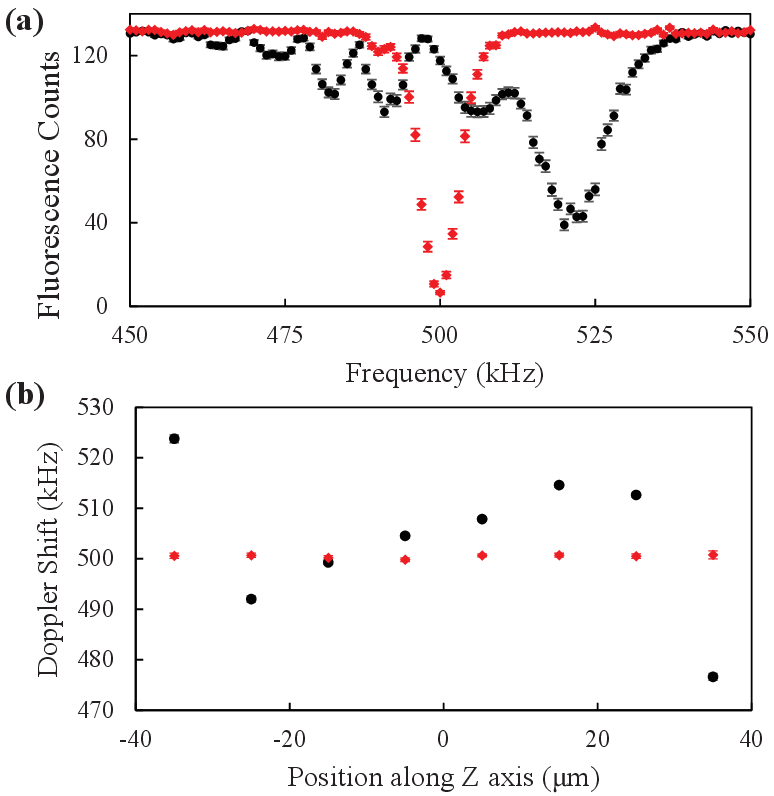}
\caption{\label{fig:Fig2} Black circles (red diamonds) indicate data taken before (after) waveform correction. (a) Frequency spectrum of the qubit transition; points represent the mean of two-ion fluorescence counts over 500 experiments, and error bars represent the standard error of the mean. (b) Doppler shift measured in each 10 $\mu$m segment of the transport region as determined from fits to spectral peaks; error bars (fit errors) are smaller than the size of the points.}
\end{figure}

Measurements of the Doppler-shifted qubit transition frequency after this procedure reveal undesired deviations in the transport velocity. We expect the frequency of the 729~nm beam to be shifted by 500~kHz for ions moving at 0.5~m/s, but initial spectroscopy of the qubit resonance realized by shuttling an ion over the entire region from L to R reveals a multi-peaked feature [black points in Fig.~\ref{fig:Fig2}(a)]. This indicates that the Doppler shift is changing during transport. We instead probe the in-flight Doppler shift in smaller 20~$\mu$s segments [see Fig.~\ref{fig:Fig1}(c)], which produces single-peaked spectra representing the local Doppler shifts plotted in Fig.~\ref{fig:Fig2}(b). The local Doppler shifts indicate velocity variations of up to 4.9\% from the expected value. To achieve a constant Doppler shift during transport, we modify the ion velocity in each segment of the waveform with an appropriate change in sampling density. When this correction is applied to each segment, the full-transit spectrum exhibits a single peak with full width 7.0 kHz at half maximum [red points in Fig.~\ref{fig:Fig2}(a)]; this width is slightly larger than the 5.6 kHz width expected for a Gaussian intensity ramp with 42~$\mu s$ $1/e^2$ time, indicating a variation of 1 kHz on a 500 kHz background Doppler shift. We refer to this corrected waveform as the ``constant-velocity waveform" below. While the MS interaction is insensitive to motional occupation to first order, motional heating of the mode during the MS gate can cause errors. We measure the motional heating after a round-trip transport using the constant-velocity waveform and observe no transport-induced heating for BM or c.m. modes (within 0.04 and 0.1 quanta uncertainty, respectively). Assuming a BM mode heating rate $\dot{\overline{n}}$ given by this uncertainty distributed over our transport duration provides an estimated upperbound error contribution from gate transport of $\frac{\dot{\overline{n}}\tau}{4} \approx 0.005$\cite{ballance_high-fidelity_2016}.

In the absence of Stark shifts, the blue and red BM motional sideband frequencies are given by $\omega_{b}=\omega_{c}+\omega_\mathrm{BM}$ and $\omega_{r}=\omega_{c}-\omega_\mathrm{BM}$, respectively, where $\omega_{c}$ is the qubit carrier frequency. When the ions are illuminated with the two MS interaction tones, the sideband frequencies experience an intensity-dependent light-shift of $\Delta_{S}$, assumed to be equal for both sidebands\cite{kirchmair_deterministic_2009, haffner_precision_2003}. To implement the entangling interaction, we apply the two MS tones at detunings (from the bare carrier) of $\delta_{b}=\omega_\mathrm{BM}+\delta_{m}+\delta_{g}$ for the blue tone and $\delta_{r}=-\omega_\mathrm{BM}-\delta_{m}+\delta_{g}$ for the red tone. Here, $\delta_m$ is the usual ``mode detuning", and $\delta_g$ is a ``global detuning" that can be used to compensate for the light shift. Both $\Delta_{S}$ and $\delta_g$ can be time-dependent.

We consider the interaction of co-propagating laser fields of equal intensity and polarization at frequency detunings $\delta_{b}$ and $\delta_{r}$ with the electric quadrupole transition of the ions \cite{james_quantum_1998}. After taking both Lamb-Dicke and rotating wave approximations and neglecting off-resonant terms, we write the MS Hamiltonian in the (Stark-shifted) spin and motion interaction frame as: 
\begin{eqnarray*}
H_\mathrm{MS} = \frac{\hslash \Omega \eta_\mathrm{BM}}{2}(a^\dagger e^{-i(\delta_{m}+\delta_{g}-\Delta_{S})t}+ a e^{i(\delta_{m}-\delta_{g}+\Delta_{S})t})\\
 \times~(\sigma^{+}_{1}-\sigma^{+}_{2}) + h.c.
\end{eqnarray*}
where $\Omega$ is the carrier Rabi frequency, $\eta_{BM} \sim 0.042$ is the BM Lamb-Dicke parameter, $a^\dagger$ and $a$ are creation and annihilation operators for BM vibrations, and $\sigma^{+}$ and $\sigma^{-}$ are raising and lowering operators for the internal electronic states of the indicated ion. We model the dynamics of the interaction by solving the time-dependent Schrodinger equation for $\tau=160$~$\mu$s, and we allow the parameters of $H_\mathrm{MS}$ to vary during the interaction to analyze the effects of a changing Rabi frequency and Stark shift in cases where the ions are transported across the 729~nm beam.

We briefly describe typical calibration procedures and results for an MS entangling gate on ions in a stationary potential in order to highlight the differences between this simpler situation and the interaction we later implement on transported ions. For a constant-intensity interaction, we perform Rabi experiments to calibrate both the red and blue BM sideband frequencies. During the calibration of each sideband, we apply the other tone at a detuning of $\delta_{m}/(2\pi)= 25$~kHz from its sideband frequency to generate the Stark shift present during the gate while only minimally driving the gate interaction. This procedure effectively applies the global detuning necessary to compensate the Stark shift ($\delta_g = \Delta_S$). We use a similar experiment to balance Rabi frequencies. Then, for a fixed bichromatic interaction duration, we analyze the ion state populations $P_0$, $P_1$, and $P_2$ as a function of $\delta_{m}$ to determine that value which minimizes $P_1$, and we adjust the beam powers at this optimum detuning to balance $P_0$ and $P_2$. For $\tau=160$~$\mu$s, we choose $\delta_{m}/(2\pi) = 12.5$~kHz (implementing two loops in motional phase space), and achieve a gate fidelity $F = 97.0(4)$\%.
This fidelity is limited largely by magnetic field noise and laser phase noise: we measure a 1.4(3)\% contrast loss in a single-ion Ramsey experiment with a 160~$\mu$s delay, suggesting an approximate decoherence-induced two-ion process error of 3\%.

\begin{figure}[htbp]
\includegraphics[height=90mm,width=0.45\textwidth]{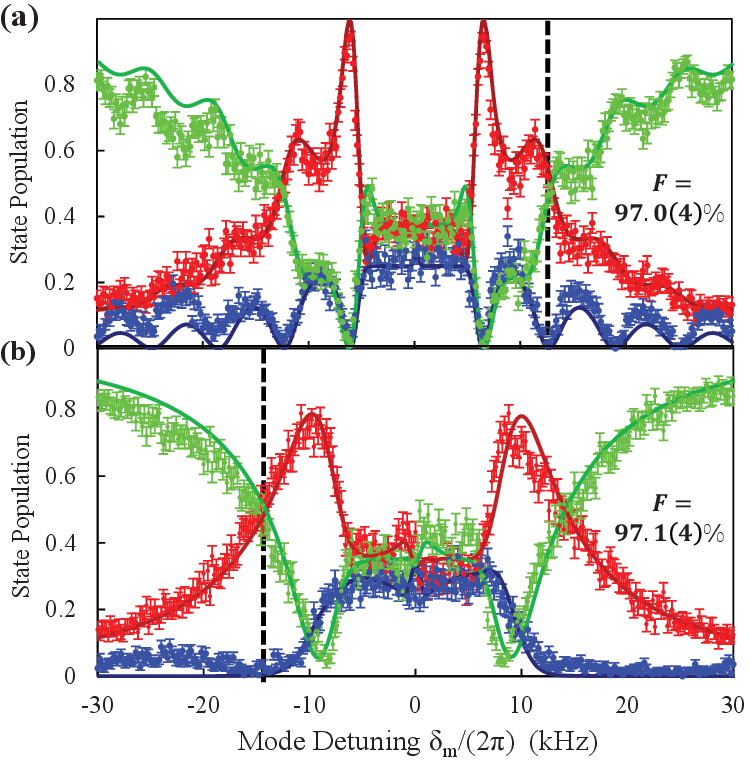}
\caption{\label{fig:Fig3} State populations $P_0$ (red), $P_1$ (blue), and $P_2$ (green) as a function of  mode detuning $\delta_m$ for $\tau =160$~$\mu$s. Points represent experimental data (error bars give the 68\% confidence  interval  assuming  binomial  statistics). Solid lines represent fits to simulations with our model Hamiltonian $H_\mathrm{MS}$ and with $\Omega$ and $\Delta_{S}$ as free parameters; in the transport gate case these are given a Gaussian time envelope. Vertical dashed lines indicate the detunings corresponding to the reported gate fidelities. (a) Constant intensity interaction with ions in a stationary potential; $F = 97.0(4)\%$. (b) Interaction on ions transported with modulated velocity through the beam; $F = 97.1(4)\%$.}
\end{figure}

When the Stark shifts are compensated ($\delta_g=\Delta_S$), the state population curves become symmetric about $\delta_m=0$, as seen in Fig.~\ref{fig:Fig3}(a). The constant-intensity square pulse of 729~nm light produces minima in $P_1$ at mode detunings $\delta_m=n/\tau$ for integers $n$, so that only these values are optimal. Revivals in $P_1$ between these optimal values can be diminished through a smooth intensity envelope \cite{leibfried_transport_2007}, thereby reducing this constraint on detunings.
When we transport ions through the stationary 729~nm Gaussian beam, they experience a natural Gaussian intensity envelope, but this smoothly varying optical intensity also creates a time-dependent Stark shift $\Delta_S(t)$, rather than a constant $\Delta_S$, that cannot be completely compensated through a constant offset in drive frequencies ($\delta_{g}$). The uncompensated Stark shift breaks the mode-detuning symmetry of the interaction, complicating the dynamics. While the light shift can be eliminated to first order by adjusting the relative tone powers \cite{kirchmair_deterministic_2009} or by illuminating the ions with an additional tone that produces an opposite light shift \cite{haffner_precision_2003}, we instead independently implement two different Stark shift compensation methods for these ``transport gates" performed on transported ions; the first ``static" method involves the application of a constant $\delta_{g}$ that only loosely counteracts the time-varying Stark shift. The second ``dynamic" method leverages in-flight adjustments to the velocity of the moving potential to smoothly counteract $\Delta_S(t)$ with a varying Doppler shift $\delta_{D}(t)$ which acts as a time-dependent $\delta_{g}(t)$. We model these effects in both cases using a Gaussian envelope on $\Omega(t)$ and $\Delta_S(t)$ that takes into account the beam waist, the moving potential, and the ion spacing.

All calibrations for the transport gate are performed with the beams turned on during the full transport duration [Fig.~\ref{fig:Fig1}(b)]. We begin by calibrating the Doppler-shifted, bare sideband frequencies ($\omega_{b}$ and $\omega_{r}$) using a reduced optical power with the constant-velocity waveform; the lowered intensity avoids over-driving the transitions and minimizes the Stark shift. We then perform the MS sequence at full power and analyze the resulting populations to determine a value for the detuning $\delta_m$ that minimizes $P_1$. We first attempt the static compensation method, for which we scan the global detuning $\delta_g$ to further minimize $P_1$ and then optimize the beam powers to balance $P_0$ and $P_2$. We perform the MS gate at $\delta_m/(2\pi)=14.2$~kHz and $\delta_g/(2\pi)=4.4$~kHz, and we measure a gate fidelity $F=96.6(4)$\%. 

The static compensation method is straightforward to implement, and it allows us to determine both the required optical power necessary to perform the transport gate and the corresponding Stark shift experienced by the ions during transport. The gate performed on transported ions requires a three-fold increase in optical intensity compared to the standard gate because the transported ions spend less time in the most intense portion of the beam. While the intensity necessary for a single gate is higher, the power could be recycled to perform gates on other ion pairs in a larger trap array\cite{leibfried_transport_2007}. To quantify the light shift variation during the transport gate, we measure the Stark-shifted sideband frequencies during 20~$\mu s$ segments of the transport at the MS gate intensity. As described above for the stationary potential well, each sideband is calibrated with the other tone applied simultaneously at $\delta_{m}/(2\pi)= 25$~kHz. We measure a 5 kHz difference between sideband frequencies at the edge of the beam and those in the most intense, center portion - a quantity consistent with the value of $\delta_g$ used to optimize gate performance.

Rather than applying a constant frequency correction for the light shift, the ion transport allows us to implement a unique dynamic Stark shift correction using our control over ion velocity. We create a ``modified-velocity waveform" by making further small adjustments to the sampling density in each segment of the waveform. The adjustments produce an additional time-varying Doppler shift $\delta_{D}(t)$ on the laser tones that counteracts the intensity-dependent Stark shift ($\delta_{g}(t)=\delta_{D}(t)=\Delta_{S}(t)$). Using the bare, Doppler-shifted sideband frequencies calibrated with the constant-velocity waveform, we implement the MS interaction on ions transported with the modified-velocity waveform. We choose a value of $\delta_m$ that minimizes $P_1$ [Fig.~\ref{fig:Fig3}(b)], and we optimize the global beam power to balance $P_0$ and $P_2$. For $\delta_m/(2\pi)=-15$~kHz, we measure a gate fidelity of $F=97.1(4)$\%. The interaction model fit to the dynamic correction transport gate data determines an uncompensated Stark shift of 140~Hz and predicts an error of 0.0058 for our gate. With perfect Stark shift compensation, the expected error reduces to 0.0026; this could be reduced even further by operating at larger gate detuning with greater optical power. The use of fine adjustments to the moving potential's velocity allows us largely to remove the symmetry-breaking light shift effects while maintaining a smooth optical intensity ramp and therefore allows a wider range of detunings and optical intensities for the gate.

While the transport used in this study is adiabatic, we note that non-adiabatic transport could also be incorporated in the gate. In that case, a well defined phase must be established between the transport and the optical force to ensure that the same motional phase-space trajectory is achieved during every experimental repetition \cite{sawyer_spin_2014-1, ruzic_entangling-gate_2021}. The subdivision of the transport waveform into eight 20 $\mu$s segments does not approach the limits of the AWG sampling rate (5 ns) nor the response time of the electrode filters (2 $\mu$s). Shorter waveform segments would provide finer velocity control for correcting Doppler shift variations and for dynamically compensating Stark shifts. Through observations of mode detuning asymmetry, we can leverage the gate itself to calibrate small corrections to the waveform interpolation which optimally compensate the variable Stark shifts. Despite the coarse discretization of the transport waveform used here, we obtain gate fidelities equalling those of stationary gates in our system and demonstrate the feasibility of a transport-based universal quantum gate set.

Ongoing advances in trapped-ion experimental control provide opportunities to explore new quantum logic architectures and techniques. Fine spatial and temporal manipulation of trap electric fields yields freedom in the confining potential position, velocity, and strength that can be exercised to modify qubit interactions in ways that can reduce constraints on experimental requirements and can improve gate performance. The transport gate technique demonstrated here could be extended to longer ion strings with different motional modes and ion species with appropriate changes in beam geometry. In particular, the scheme is amenable to $\sigma^z\sigma^z$ interactions such as the optical-transition dipole force gate \cite{sawyer_wavelength-insensitive_2021, clark_high-fidelity_2021-1} where the Stark shift effects could be eliminated entirely with an echo pulse.

\begin{acknowledgments}
Research was sponsored by the Army Research Office and was accomplished under Grant Number W911NF-18-1-0166. The views and conclusions contained in this document are those of the authors and should not be interpreted as representing official policies, either expressed or implied, of the Army Research Office or the U.S. Government. The U.S. Government is authorized to reproduce and distribute reprints for Government purposes notwithstanding any copyright notation herein.
\end{acknowledgments}

\bibliography{TransportGateBib}

\end{document}